\documentclass[reprint, showpacs, preprintnumbers, nofootinbib, amsmath, amssymb, aps, prd, floatfix, groupedaddress]{revtex4-1}

\usepackage{url}
\usepackage{xcolor}
\usepackage{amsmath}
\usepackage{amssymb}
\usepackage{graphicx}
\usepackage{algorithm,algpseudocode}
\usepackage[frozencache,cachedir=./]{minted}
\usepackage{mathrsfs}
\usepackage{microtype}
\usepackage{aas_macros}
\usepackage{cancel}
\usepackage[normalem]{ulem}

\makeatletter
\renewcommand{\fnum@algorithm}{\fname@algorithm}
\makeatother

\newcommand{\bham}{School of Physics \& Astronomy and Institute for Gravitational Wave Astronomy, University of Birmingham, Birmingham, B15 2TT, UK}

\begin{document}

\title[Label Switching Problem]{The Label Switching Problem in Bayesian Analysis for Gravitational Wave Astronomy}

\author{Riccardo Buscicchio}
\email{riccardo@star.sr.bham.ac.uk}

\author{Elinore Roebber}

\author{Janna M. Goldstein}

\author{Christopher J. Moore}

\affiliation{\bham}

\date{\today}

\begin{abstract}
The label switching problem arises in the Bayesian analysis of models containing multiple indistinguishable parameters with arbitrary ordering. 
Any permutation of these parameters is equivalent, therefore models with many such parameters have extremely multi-modal posterior distributions.
It is difficult to sample efficiently from such posteriors.
This paper discusses a solution to this problem which involves carefully mapping the input parameter space to a high dimensional \emph{hypertriangle}.
It is demonstrated that this solution is efficient even for large numbers of parameters and can be easily applied alongside any stochastic sampling algorithm.
This method is illustrated using two example problems from the field of gravitational wave astronomy.
\end{abstract}
\maketitle
\section{Introduction} \label{sec:1}
It sometimes occurs in Bayesian inference problems that the target distribution depends on several parameters whose ordering is arbitrary.
Three examples are immediately apparent from the field of gravitational wave (GW) astronomy alone.
Firstly, when describing a compact binary with component masses $m_1$ and $m_2$, the likelihood is symmetric under exchange of the labels 1 and 2 (provided all other relevant parameters are suitably adjusted simultaneously).
Secondly, when analysing GW time series data containing two or more overlapping sources of the same type, the likelihood is invariant under exchanging all of the parameters of any pair of sources.
And thirdly, when analysing the parameters of a population of observed GW events, mixture models can be used to model the population and/or to infer the presence of distinct astrophysical populations. In this case the hyper-likelihood for the population parameters may be invariant under exchanging the parameters of the population components.

Sometimes a simple reparametrisation and restricting the parameter range is enough to remove the degeneracy arising from the arbitrary ordering. 
In the first case of the binary with two component masses, it is possible to define, say, the total mass $M=m_1+m_2$ and mass ratio $q = m_2/m_1$ and to sample these over the restricted ranges $M>0$ and $q\leq 1$.
This covers only the restricted portion of the parameter space $m_1 \geq m_2$, thereby removing the symmetry from the likelihood.

The second and third examples are more problematic as they are not restricted to just 2 degrees of freedom.
In each case the target distribution has a high degree of symmetry and is invariant under permutations of some number of labels, $K$.
A great deal of literature is devoted to this \emph{label switching problem} in the context of mixture models \cite{10.2307/2030064, 10.2307/2345907, doi:10.1111/1467-9868.00095, doi:10.1111/1467-9868.00265, jasra2005, doi:10.1198/1061860031329, 10.2307/2670359}.
The invariance of the target distribution under permutations means that if the posterior has a peak (or mode) at a particular point in parameter space it will necessarily have peaks at all $K!$ points related by symmetry.
The extreme scaling of this multimodality poses a serious obstacle to any sampling algorithm in moderate or high dimensional problems.

The most natural way to solve the label switching problem is to impose an \emph{artificial identifiability constraint}.
Searching over the restricted region $m_1 \geq m_2$ of the binary component mass space is an example of such a constraint in 2 dimensions.
In the $K$ dimensional problem this can be generalised by demanding a certain ordering of the parameters; see, for example, \cite{10.2307/2345907, doi:10.1111/1467-9868.00265, jasra2005, doi:10.1198/1061860031329}.
Restricting to this small region of parameter space avoids all symmetries and removes the excess multimodality.
It is also obvious that if one can adequately explore the restricted parameter space satisfying the artificial identifiability constraint then, by symmetry, this is equivalent to exploring the full space.

It remains to implement a suitable artificial identifiability constraint in practical inference problems.
This problem can be approached in several ways. 
For example, when using an Markov chain Monte Carlo (MCMC) to explore the target distribution the proposal can be augmented by composing with a sorting function; i.e. propose a point then reorder the parameters such that the constraint is satisfied \cite{10.2307/2670359}.
Alternatively, the log-prior distribution can be crudely modified so that it returns $-\infty$ for any point not satisfying the constraint.
Either of these will ensure the chain never leaves the desired region of parameter space.

While undoubtedly simple, neither of these approaches are completely satisfactory. The former approach requires the user to modify their MCMC proposal distribution and it is difficult to apply when using other stochastic sampling algorithms which don't have a user-accessible proposal distribution (such as nested sampling \cite{2004AIPC..735..395S}).
For this reasons such an approach is not compatible with the modern approach of treating the sampler, as far as possible, as a \emph{black box} to which the user must only provide a likelihood and a prior.
The latter approach is easy to implement for all samplers, but has the significant drawback of being extremely inefficient in high numbers of dimensions.
This is because the sampler only proposes useful points satisfying the identifiability constraint a tiny fraction $1/K!$ of the time.

This paper presents a solution to the label switching problem. 
Our approach follows that of \cite{2015MNRAS.453.4384H} (see, in particular, Eq.~A13; however, this equation contains a typographical error as pointed out by \cite{2019MNRAS.485.5363G}).
This solution to the label switching problem has been implemented in \cite{polychord_lite} and has been widely used extensively in the astronomical and cosmological literature \cite{2019arXiv190800906H, 2019MNRAS.485.5363G, 2019MNRAS.483.4828H, 2018arXiv180706211P, 2018MNRAS.479.2968H, 2018A&A...617A..96M, 2018MNRAS.481.3853O, 2018JCAP...04..016F, 2017MNRAS.466..369H, 2016A&A...594A..20P, 2016MNRAS.455.2461H}.
Here we describe the solution in detail, including proofs of certain important properties of the solution.
This solution is mathematically elegant, efficient in high dimensions, and can be easily integrated with any sampling algorithm while treating it as a black box.

In Sec.~\ref{sec:2} the label switching problem is described in detail and the idea behind our proposed solution is illustrated in 2 dimensions. 
Our solution, for an arbitrary number of dimensions, is presented in Sec.~\ref{sec:3}.
In Sec.~\ref{sec:4} the efficacy of our proposed solution is demonstrated by applying it to the second (Sec.~\ref{subsec:4a}) and third (Sec.~\ref{subsec:4b}) example problems described in the opening paragraph of this section.
These example applications are drawn from the field of GW astronomy, but we stress that this method has been more generally applied to inference in astronomy already.

\section{The Label Switching Problem}\label{sec:2}

We wish to treat problems containing multiple indistinguishable components.
Each of the $K$ components is modeled by some parameters $\mathbf{\Lambda}_k\in U$, where the parameter space $U$ is an open set of $\mathbb{R}^{n}$ and $k\in\{1,2,\ldots,K\}$. 
We will choose to distinguish components based on the values of one of these parameters, $x_k\equiv\mathbf{\Lambda}_k^1\in I$ where $I$ is an set of $\mathbb{R}$.
For simplicity, in this section we will consider $x_k \in(0,1)$ and use a flat prior on each $x_k$, although these restrictions can be relaxed later. 

In the case where there are two components, $K=2$, our full parameter space is $U \times U$.
However we will mainly be interested in the subspace spanned by $\vec{x}=(x_1, x_2)$, which covers the unit square $I\times I$ (in the general $K$-dimensional case this will be a hypercube which will be denoted $\mathcal{C}$).
For the remainder of this section we suppress the other components of $\mathbf{\Lambda}_k$ from our notation for clarity.

Since the two components are indistinguishable, the points $(x_1,x_2)$ and $(x_2,x_1)$ are equivalent; both the likelihood, $\mathcal{L}(\vec{x})$, and prior distributions are symmetric under interchange of the labels 1 and 2 (provided we also remember to relabel all the other components of $\mathbf{\Lambda}_1$ and $\mathbf{\Lambda}_2$ simultaneously). 
As a result, the parameter space is twice as large as it needs to be.
Evaluating $\mathcal{L}(\vec{x})$ over the square will typically lead to a distribution with two global maxima (an exception occurs when the true maximum is on the boundary $x_1=x_2$); secondary peaks, ridges and other structures in the likelihood are also duplicated.
In higher dimensions this duplication and multimodality increases in proportion to $K!$ and becomes a serious obstacle to sampling the target distribution. 

\begin{figure}[ht]
    \centering
    \includegraphics[width=\columnwidth]{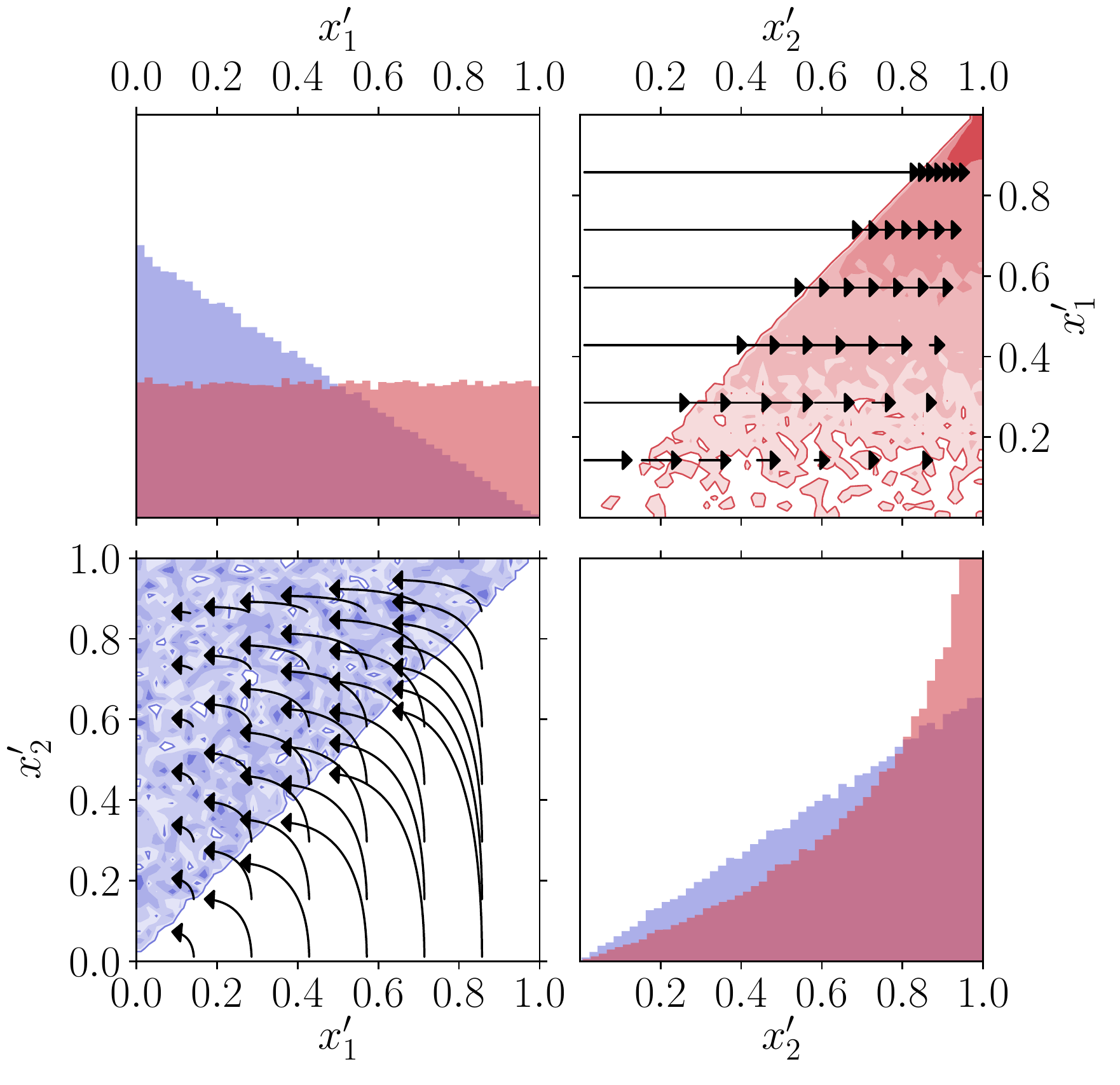}
    \caption{
    Two overlaid corner plots, one in the lower-left triangle (blue) and the other in the upper-right triangle (red).
    Points $\vec{x}=(x_1,x_2)$ were drawn uniformly in the unit square $10^{5}$ times.
    Histograms of the points $\vec{x}'=\phi(\vec{x})$ are plotted for both the map in Eq.~\ref{eq:mapbad} (red) and the map in Eq.~\ref{eq:mapgood} (blue).
    Both maps correctly move points from the square to the triangle, but only Eq.~\ref{eq:mapgood} does so while preserving the correct uniform prior.
    The arrows illustrate how points in the square move under the action of the two maps. 
    }
    \label{fig:cornerplots}
\end{figure}

To avoid sampling multiple identical copies of the same likelihood modes we will enforce the identifiability constraint $x_2\geq x_1$. 
This amounts to labelling the component with the smallest $x$ as \#1, the component with the next largest $x$ as \#2, and so on in higher dimensions. 
In two dimensions, this restricts the parameter space to the triangle $x_2\geq x_1$ (see the off principal diagonal panels in Fig.~\ref{fig:cornerplots}).  In higher dimensions, the parameter space is restricted to the region $x_K\geq x_{K-1}\geq \cdots \geq x_1$, which is hereafter referred to as the hypertriangle and denoted $\mathcal{T}$.

Samplers naturally propose points in a hypercube. To avoid modifying the sampler itself, we wish to map points in the hypercube to points in the hypertriangle (following the strategy first introduced in the astronomy literature by \cite{2009MNRAS.398.1601F}):

\begin{align}
    \phi:\mathcal{C}\rightarrow\mathcal{T}\,.
\end{align}

Naively, we might try to choose $\phi$ to be the sorting function. 
Unfortunately, although it does map into the hypertriangle, it doesn't solve the multimodality problem, since sorting is a many-to-one map. If the sampler proposes a point $\vec{x}=(x_1,x_2)$ in the hypercube and then the user applies the sorting function $\vec{x}'=\mathrm{sort}(\vec{x})$ before evaluating the likelihood $\mathcal{L}(\vec{x}')$, nothing restricts the sampler from searching over the full hypercube. 
In fact, this procedure is identical to sampling the original hypercube with no sorting.

This is to be distinguished from the procedure of sorting inside the proposal distribution, as referenced in the introduction, which does restrict sampling to the hypertriangle. 
This is because the newly sorted points are kept by the sampler and used for generating the next set of proposed points. 
However, this approach violates our desire to treat samplers as black boxes.

To solve the problem we seek a function, $\phi$, which is one-to-one. 
One possibility, in 2 dimensions, is to leave the $x_1$ coordinate invariant and shift/rescale the $x_2$ coordinate such that it lies in the desired range: 
\begin{align}
    x'_{1} &= x_{1} \label{eq:mapbad} \\ 
    x'_{2} &= x'_{1}+(1-x'_{1})x_{2} \,. \nonumber
\end{align}
To see that points are indeed mapped to $\mathcal{T}$ it is sufficient to note that the correct ordering is enforced by adding a positive quantity to $x'_1$ to get a larger value for $x'_2$. 
The range $x'_2\in(0, 1)$ is in turn ensured by scaling $x_2$ with the factor $1 - x'_1$.
This map is a indeed one-to-one map from the square to the triangle, thereby removing the problem of multiple modes. 
However, this map has the unfortunate property that it distorts the prior on the $x_2$ component, favoring larger values (see the red distribution in Fig.~\ref{fig:cornerplots}).

The map in Eq.~\ref{eq:mapbad} can be ``fixed'' by revising the $x_1$ coordinate downwards, before shifting/rescaling $x_2$:
\begin{align}
    x'_{1} &= 1-\sqrt{1-x_{1}} \label{eq:mapgood} \\
    x'_{2} &= x'_{1}+(1-x'_{1})x_{2} \,. \nonumber
\end{align}
The new map in Eq.~\ref{eq:mapgood} solves the problem in 2 dimensions (see the blue distribution in Fig.~\ref{fig:cornerplots}). 
The sampler can propose points $\vec{x}=(x_1,x_2)$ uniformly in the square. These points are mapped to the triangle $\vec{x}'=\phi(\vec{x})$. Finally, the likelihood is evaluated at the mapped points, $\mathcal{L}(\vec{x}')$. 
This procedure correctly covers the parameter space just once with the desired flat prior.
To prove that the proposed map does indeed maintain the desired flat prior on the individual components one can evaluate the Jacobian of the transformation $\vec{x}'=\phi(\vec{x})$ and show that is constant. This is done in the next section for the $K$-dimensional case. 
Because the Jacobian is constant, this transformation will correctly preserve the flat prior that is imposed on the original $x_k$.~\footnote{For an extension of our solution to the wider class of separable priors, see Sec.~\ref{subsec:3c}.}

To state the problem formally: to solve the label-switching problem, we seek a bijection (a ``one-to-one'' and ``onto'' map) ${\phi:\mathcal{C}\rightarrow\mathcal{T}}$, for an arbitrary number of dimensions, with components ${x'_{\kappa}=\phi_{\kappa}(x_k)}$, such that the Jacobian $J\equiv\left|\partial x'_{\kappa} / \partial x_{k} \right|=\mathrm{const}$.
\section{The hypertriangle map in arbitrary dimensions}\label{sec:3}
Our proposed generalization of the 2-dimensional map in Eq.\ref{eq:mapgood},  $x'=\phi(x)$,  is defined recursively as
\begin{align} \label{eq:map}
    x'_i = x'_{i-1}+(1-x'_{i-1})\left[1-(1-x_i)^{\frac{1}{K+1-i}}\right] \,,
\end{align}
where $i\in {1,\ldots,K}$ and $x'_{0}= 0$ by definition. This closely resembles Equation (A13) of \cite{2015MNRAS.453.4384H}, although here we have corrected a typographical error. Eq.~\ref{eq:map} can be expressed non-recursively as:
\begin{equation} \label{eq:map2}
    x'_i = 1 - \prod_{j=1}^{i} (1 - x_j)^{\frac{1}{K+1-j}} \;.
\end{equation}
If the inputs are in the correct range $x_j\in(0,1)$, i.e. $x_j\in\mathcal{C}$, it can be shown that the output falls in $\mathcal{T}$ (the logic as outlined in Sec. \ref{sec:2} for Eq. \ref{eq:mapbad} still applies). It can also be shown that this map is a bijection by inverting Eq. \ref{eq:map2}.

In the remainder of the section, we will first prove that Eq.\ref{eq:map2} is equivalent to Eq.\ref{eq:map}, and then that the Jacobian of Eq.\ref{eq:map2} is constant. 

\subsection{Equivalent Representations of $\phi$}

Starting with the recursive version of the map given in Eq.\ref{eq:map}, we rearrange it as follows:
\begin{align}
x_{i}^{\prime} & =  x_{i-1}^{\prime}+\left(1-x_{i-1}^{\prime}\right)\left[1-\left(1-x_{i}\right)^{\frac{1}{K+1-i}}\right]\nonumber\\
 & =  \left[1-\left(1-x_{i}\right)^{\frac{1}{K+1-i}}\right]-x_{i-1}^{\prime}\left[1-\left(1-x_{i}\right)^{\frac{1}{K+1-i}}-1\right]\nonumber\\
 & =  \left[1-\left(1-x_{i}\right)^{\frac{1}{K+1-i}}\right]+\left(1-x_{i}\right)^{\frac{1}{K+1-i}}x_{i-1}^{\prime}\nonumber\\
  & =  1-\left(1-x_{i}\right)^{\frac{1}{K+1-i}}\left(1-x_{i-1}^{\prime}\right)\,.
\end{align}
This procedure can be repeated for the $x_{i-1}^\prime$ term inside the final set of parentheses, and then for $x_{i-2}^\prime$ and so on down to $x^\prime_1$.
This gives the equivalent representation to Eq.~\ref{eq:map2};
\begin{align}
x_{i}^\prime & =  1-\left(1-x_{i}\right)^{\frac{1}{K+1-i}}\bigg[\left(1-x_{i-1}\right)^{\frac{1}{K+1-(i-1)}}\left(1-x_{i-2}^{\prime}\right)\bigg]\nonumber\\
 & = \ldots \nonumber \\
 & =  1-\prod_{j=1}^{i}\left(1-x_{j}\right)^{\frac{1}{K+1-j}} \;.
\end{align}

\subsection{The Jacobian of $\phi$}

As discussed in Sec.~\ref{sec:2}, to maintain the correct prior on the hypertriangle, it is necessary that the map $\phi$ has a constant Jacobian.  
To prove that our proposed hypertriangle map has this property, we start with the form of the map in Eq.\ref{eq:map2}.
The Jacobian matrix for this specific transformation is lower-triangular because the component $x'_i$ depends only on $x_j$ with $j\leq i$. Its determinant is therefore equal to the product of the diagonal terms:
\begin{eqnarray}
J & = & \prod_{i=1}^{K}\frac{\partial x_{i}^{\prime}}{\partial x_{i}}  \nonumber\\
  & = & \prod_{i=1}^{K}\frac{1}{K+1-i}\left(1-x_{i}\right)^{\frac{1}{K+1-i}-1}\prod_{j=1}^{i-1}\left(1-x_{j}\right)^{\frac{1}{K+1-j}}\nonumber \\
  & = & \frac{1}{K!}\prod_{i=1}^{K}\frac{1}{(1-x_{i})} \left(1-x_{i}\right)^{\frac{1}{K+1-i}}\prod_{j=1}^{i-1}\left(1-x_{j}\right)^{\frac{1}{K+1-j}} \nonumber \\
 & = & \frac{1}{K!}\prod_{i=1}^{K}\frac{1}{(1-x_{i})}\prod_{j=1}^{i}\left(1-x_{j}\right)^{\frac{1}{K+1-j}}  \label{eq:old} \,, 
\end{eqnarray}
where in the final step a factor has been moved inside of the second product and the upper limit of the product has been changed accordingly.
Writing out the products explicitly gives 
\begin{align}
 J&= \frac{1}{K!} \nonumber \\
 &\,\times
\frac{1}{\left(1-x_{1}\right)}\left[\left(1-x_{1}\right)^{\frac{1}{K+1-1}}\right]\nonumber\\
 & \,\times\frac{1}{\left(1-x_{2}\right)}\left[\left(1-x_{1}\right)^{\frac{1}{K+1-1}}\left(1-x_{2}\right)^{\frac{1}{K+1-2}}\right]\nonumber\\
 & \,\times\dots\nonumber\\
 & \,\times\frac{1}{\left(1-x_{K}\right)}\left[\left(1-x_{1}\right)^{\frac{1}{K+1-1}}\dots\left(1-x_{K}\right)^{\frac{1}{K+1-K}}\right]
\end{align}
Careful counting of all the terms reveals that everything cancels and we are left with
\begin{align} \label{eq:Jac_phi}
 J = \frac{1}{K!}\,.
\end{align}
The Jacobian is equal to one over the number of times the original parameter space was covered by the hypercube.
\subsection{Extension to separable priors}\label{subsec:3c}
The above derivation considered only flat priors on the $x_k$.
Here we consider the applicability of our hypertriangulation map to separable priors of the form 
\begin{equation}
\Pi(x_1,\dots,x_K)=\prod_{k=1}^K\pi(x_k)\,
\end{equation}
In such cases it is first necessary to transform to new coordinates such that the prior is flat before proceeding to apply the hypertriangulation map as before.

In order to find the new coordinates with flat priors, first evaluate the cumulative distribution function
\begin{equation}
F(x)=\int_{0}^{x}\pi(s)\text{d}s\,.
\end{equation}
Then define new coordinates $y_k=F(x_k)$ which lie in the range $[0,1]$.
The prior on these new coordinates is now flat and the hypertriangulation map may now be applied to the $y_k$.

\subsection{Implementation of $\phi$}
\label{subsec:3d}

For concreteness, we provide here a pseudo-Python implementation of Eq.~\ref{eq:map2}.
The input \texttt{x} (in $\mathcal{C}$) and output \texttt{X} (in $\mathcal{T}$) are numpy arrays where all values are in the prior range $(0,1)$. 
The values of \texttt{x} may be in any order whilst the values of \texttt{X} are, by construction, in ascending order.
If a different prior range is needed then the input and output must be shifted and rescaled as appropriate.
A full Python implementation (including the shifting and rescaling) is provided at the GitHub repository~\cite{python_script}. 
\begin{algorithm}[H]
  \caption{Python Implementation of Eq.~\ref{eq:map2}}
  \begin{algorithmic}[1]
    \State \texttt{def phi(x):}
    \State\hspace{0.3cm} \texttt{K = len(x)}
    \State\hspace{0.3cm} \texttt{i = numpy.arange(K)}
    \State\hspace{0.3cm} \texttt{inner = numpy.power(1 - x, 1/(K - i))}
    \State\hspace{0.3cm} \texttt{X = 1 - numpy.cumprod(inner)}
    \State\hspace{0.3cm} \texttt{return X}
  \end{algorithmic}
\end{algorithm}
\section{Example GW Applications}\label{sec:4}

In this section we present two applications of our hypertriangle method to two rather different Bayesian inference problems drawn from the field of GW astronomy.

The first example in Sec.~\ref{subsec:4a} is a Gaussian mixture model; models of this type have been studied extensively in the context of the label switching problem \cite{doi:10.1198/1061860031329, doi:10.1111/1467-9868.00095, jasra2005, 10.2307/2670359, doi:10.1111/1467-9868.00265, 10.2307/2345907, 10.2307/2030064}. 

The second example in Sec.~\ref{subsec:4b} involves the identification of multiple overlapping signals in time series data. The label switching problem has not often been explicitly considered in this context. 
However \cite{article_multi_damped_sin} discuss it when fitting multiple damped sinusoids to time series data. 

\subsection{The Observed Mass Function of LIGO/Virgo Binary Black Holes}\label{subsec:4a}

LIGO and Virgo \cite{2015CQGra..32g4001L, 2015CQGra..32b4001A}
are ground-based GW detectors operating in the $(10^{1}-10^{4})\,\mathrm{Hz}$
frequency range. 
The network has been operating since September 2015 and has so far confidently detected 10 binary black hole (BBH) mergers and 1 binary neutron star merger \cite{2018arXiv181112907T}. The third observation run is ongoing and low latency pipelines \cite{2019arXiv190108580S, 2016CQGra..33u5004U, 2012PhRvD..86b4012H, 2016CQGra..33q5012A, 2008CQGra..25k4029K} have produced a number of public alerts associated with event candidates \cite{GraceDB, Chirp}.
It is likely that by the end of the current run dozens more detections will be available \cite{2018LRR....21....3A}
for further investigation.

Detailed waveform models for BBH signal calibrated against numerical relativity are now available  \cite{2017PhRvD..95d4028B,2016PhRvD..93d4006H,2016PhRvD..93d4007K,2014PhRvL.113o1101H,2017arXiv170501845D}.
These are used in the \textsc{LALInference} Bayesian analysis software package \cite{2015PhRvD..91d2003V} 
to construct posterior distributions on the parameters of each event.
These include both intrinsic (component masses, spins, angular momentum, etc) and extrinsic (sky position, distance, inclination) parameters.

Of these parameters, the best measured and most astrophysically interesting are the individual black hole masses.
Parameter reconstruction is crucial from an astrophysical perspective, because it allows both for in-depth studies of individual objects \cite{2018arXiv181112907T,2018arXiv181100364T,2019PhRvX...9a1001A,2019arXiv190304467T, 2019arXiv190500869I,2017Natur.551...85A} and of populations masses \cite{2018arXiv181112940T,2019arXiv190306881K,2018ApJ...856..173T}.

From a statistical point of view, Bayesian inference on a population of events with imperfect measurements has a well established formalism \cite{2019MNRAS.486.1086M,2010ApJ...725.2166H}.
A residual freedom remains in the choice of parameterization for the population. 
Previous studies have used astrophysically motivated functional dependencies \cite{2018arXiv181112940T,2018ApJ...856..173T,2019arXiv190306881K,2017arXiv171202643W,2018arXiv180506442W,2017MNRAS.471.2801S}.
For example, one parameter in such models might be the location of a mass gap in the black hole population \cite{2019MNRAS.484.4216R, 2018arXiv180204909B}.
Other studies have used a broader family of somewhat non-parametric models \cite{2018MNRAS.479..601D,2019arXiv190504825P,2019MNRAS.485.1665K,hjort_holmes_müller_walker_2010,OrbTeh2010a}. 

Within the latter formalism, greater flexibility can be achieved by fitting the observed data with an unknown number of sub-components. 
No \emph{a priori} physical meaning is necessarily associated with these components, and they are usually sampled from a common hyper-parameter space.
The lack of any hierarchy among these components naturally introduces a symmetry under permutations and leads to the label switching problem.

Here we apply our hypertriangle approach to inference on the population of observed BBH component masses, $m_{1}\geq m_{2}$. 
We model the observed distribution of \emph{source frame} \cite{2016PhRvL.116f1102A} component black hole masses (in solar mass units) as a mixture $p_\mathrm{pop}(\log m_1,\log m_2)$ of $K$ bivariate Gaussians; 
\begin{align}
    \begin{bmatrix}
            \log m_1 \\ \log m_2
          \end{bmatrix}
          \sim \sum^K_{k=1} w_k\, \mathcal{N}\left(
           \begin{bmatrix}
            \mu_k^{(\log m_1)} \\ \mu_k^{(\log m_2)}
          \end{bmatrix},
           \mathbf{\Sigma}_k
          \right) \,.
\end{align}
Each component has a pair of means, $\mu^{(\log m_1)}_{k}$ and $\mu^{(\log m_2)}_{k}$, a symmetric $2\times 2$ covariance matrix, $\mathbf{\Sigma}_{k}$, and a weight, $w_{k}$. 
The covariance matrix is described by its two eigenvalues, $\lambda^1_k$ and $\lambda^2_k$, and a rotation angle $\phi_k$. 
Overall, each component is fully described by the parameter vector
\begin{align}
\boldsymbol{\Lambda}_{k}=\left(\mu^{(\log m_1)}_k,\mu^{(\log m_2)}_k, \lambda^1_k, \lambda^2_k, \phi_k, w_k \right)\,. \label{eq:LIGO_comp}
\end{align}

We choose to enforce the artificial identifiability constraint $\mu_{k+1}^{(\log m_1)}\geq\mu_{k}^{(\log m_1)}$. 
This is done by applying our map $\phi$ from Eq.~\ref{eq:map2} to the vector of components $\mu_k^{(m_1)}$ with $k=1,2,\ldots,K$.
We can sample on the modified parameter space covered by
\begin{align}
\boldsymbol{\Lambda}_{k}=\left(\chi_k,\mu^{(\log m_2)}_k, \lambda^1_k, \lambda^2_k, \phi_k, w_k \right)\,. \label{eq:LIGO_comp_2}
\end{align}
where $\mu^{(m_1)}_k=\phi(\chi_k)$. 
In the language of Sec.~\ref{sec:2}, sampling on the parameter space in Eq.~\ref{eq:LIGO_comp} covers $\mathcal{C}$ (with multimodality) while sampling on Eq.~\ref{eq:LIGO_comp_2} covers $\mathcal{T}$.

The priors are taken to be flat on all of the components in Eqs.~\ref{eq:LIGO_comp} and \ref{eq:LIGO_comp_2}, except for the $\lambda^1_k,\lambda^2_k$ which we take log-uniformly distributed within their ranges.

The ranges for $\chi_k$, $\mu_{k}^{(\log m_1)}$, $\mu_{k}^{(\log m_2)}$ are $(0,2)$, with the additional constraint of $\mu_{k}^{(\log m_1)}>\mu_{k}^{(\log m_2)}$.
The range on the angle $\phi_k$ is $(0,\pi/2)$ and the ranges on $\lambda_k^1$ and $\lambda_k^2$ are $(0.01, 4)$.
Finally, the weights $w_k$ were sampled in the range $(0,1)$ and then normalized such that $\sum_k w_k=1$.

\begin{figure}[t]
    \centering
    \includegraphics[width=\columnwidth]{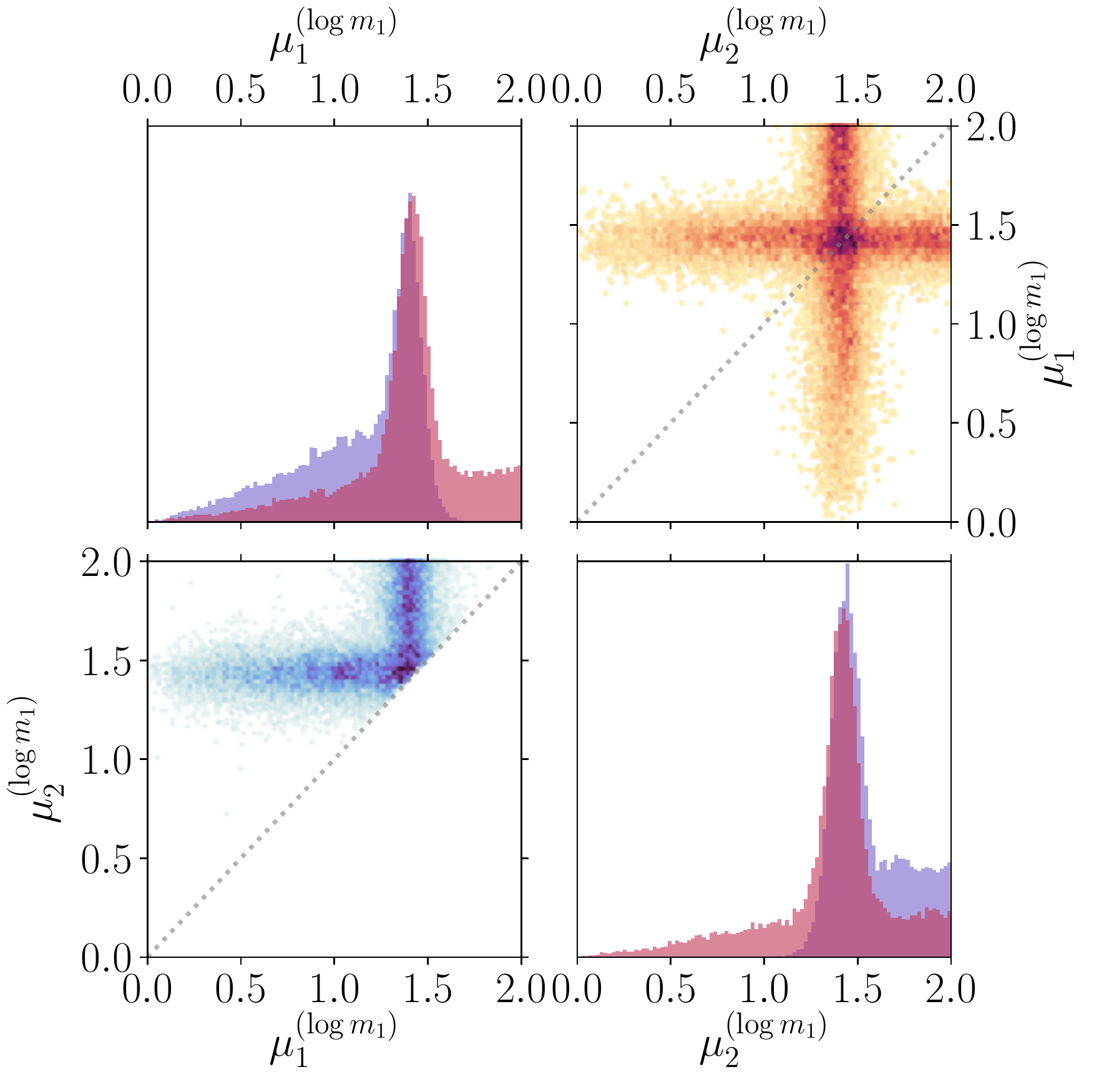}
    \caption{
    Two overlaid corner plots, one in the lower-left triangle (blue) and the other in the upper-right triangle (red).
    The red posterior is obtained by sampling in the parameter space of Eq.~\ref{eq:LIGO_comp}; this space covers the hypercube $\mathcal{C}$ and has a multimodal posterior.
    The blue posterior is obtained by sampling in the parameter space of Eq.~\ref{eq:LIGO_comp_2} and then transforming to $\mu_k^{(m1)}=\phi(\chi_k)$; this only covers the hypertriangle $\mathcal{T}$ and has a single posterior mode.
    The grey dotted line marks equality between the two components.}
    \label{fig:mix_HT_HC}
\end{figure}

We adopt a fully Bayesian hierarchical approach.
At the lowest level there are the short segments of time series data $\{d\}$ surrounding each of the $N_\text{obs}$ events.
Each event is described by some parameters $\theta$ (e.g.\ masses, spins, etc).
The likelihood that we wish to sample from is the probability of all the observed data given a certain value of the population parameters $\boldsymbol{\Lambda}=\{\boldsymbol{\Lambda}_k|k=1,2,\ldots,K\}$:
\begin{align}
p\left(\{d\}\mid \boldsymbol{\Lambda}\right)= \prod_{i=1}^{N_\text{obs}} \frac{\int \mathrm{d}\theta\, p(d \mid \theta)\,p_{\text{pop}}(\theta \mid \boldsymbol\Lambda)}{\int \mathrm{d}\theta\, p_{\text{pop}}(\theta \mid \boldsymbol\Lambda)}\,.
\end{align}
Using Bayes theorem, the above likelihood can be turned into a posterior on the population parameters $\mathbf{\Lambda}$.
This in turn can be expressed in terms of the $N_i$ posterior samples on $\theta$ from each individual event \cite{2019MNRAS.486.1086M}:
\begin{align}
p\left(\boldsymbol\Lambda \mid {\{{d}\}}\right) = \varpi(\boldsymbol\Lambda)\prod_{i=1}^{N_{\text{obs}}}\dfrac{\frac{1}{N_i}\sum_{j=1}^{N_i}\dfrac{p_{\text{pop}}(\theta_{i}^j \mid \boldsymbol\Lambda)}{\pi(\theta_{i}^j)}}{\int \mathrm{d}\theta\, p_{\text{pop}}(\theta \mid \boldsymbol\Lambda)}, \label{eq:pop_post}
\end{align}
where the posterior samples for each event are denoted $\theta_i^j$ ($i$ labels the event and $j$ labels the sample in the posterior chain), and $\varpi(\boldsymbol\Lambda)$ and $\pi(\theta)$ respectively denote the priors on the population and individual event parameters. 
We will consider only the component masses as event parameters, $\theta=(m_1,m_2)$. 
Note that the normalization integral in the denominator of Eq.~\ref{eq:pop_post} is evaluated over the constrained prior range $\log m_1 > \log m_2$.
We use the publicly available posterior samples \cite{Vallisneri_2015} for the 10 BBH events described in \cite{2018arXiv181112907T}. 

\begin{table}[t]
        \begin{center}
            \begin{tabular}{ccc}
                $\quad K$ &$ \log Z_\mathcal{T} $ & $ \log Z_\mathcal{C}$ \\
                \hline
                $\quad1$ & \multicolumn{2}{c}{ $-74.76 \pm 0.09\quad$}\\
                $\quad2$ & $\quad -78.37 \pm 0.05\quad$ & $-78.30 \pm 0.05\quad$\\
                $\quad3$ & $\quad -81.82 \pm 0.09 \quad$ & $-81.66 \pm 0.08\quad$\\
                $\quad4$ & $\quad -84.58 \pm 0.06 \quad$ & $-84.2 \pm 0.1\quad$\\
                \end{tabular}
        \end{center}
        \caption{\label{tab:logZ_LIGO}
        Log-evidences for mixtures with different number of components $K$. 
        The variables $Z_\mathcal{T}$ and $Z_\mathcal{C}$ denote the evidences obtained by sampling on the hypertriangle parameter space in Eq.~\ref{eq:LIGO_comp} and the (multimodal) hypercube parameter space in Eq.~\ref{eq:LIGO_comp_2} respectively.
        Mathematically we have already proved that these parameter spaces are equivalent and therefore the two evidences are equal; these two columns serve to demonstrate this numerically.
        For the $K=1$ component case there is no distinction between the two parameter spaces (the map $\phi$ reduces to the identity in this case).
        The errors on the CPNest evidence integrals were estimated by a combination of the internal CPNest error estimate (as described in \cite{2004AIPC..735..395S}) and examination of the spread of results from multiple runs.
        The $Z_\mathcal{T}$ and $Z_\mathcal{C}$ evidences are broadly consistent; however for large $K$ there is some tension. We think this is due to CPNest systematically underestimating the $Z_\mathcal{C}$ evidence which comes from a high dimensional and highly multimodal posterior. 
        Alternative nested samplers \cite{2009MNRAS.398.1601F, 2015MNRAS.453.4384H, 2016arXiv160603757B, 2019arXiv190402180S} have been shown to reliably estimate evidences for problems of similar complexity.}
\end{table}

As our focus here is on the label switching problem, and its solution using the  hypertriangle map, for simplicity we do not consider selection effects \cite{2014ApJ...795...64F,2014arXiv1412.4849F}.
Rather, we model the distribution of \emph{observed} black hole masses.
We defer a full treatment, including selection effects, to future work.

\begin{figure*}[t]
\includegraphics[width=2\columnwidth]{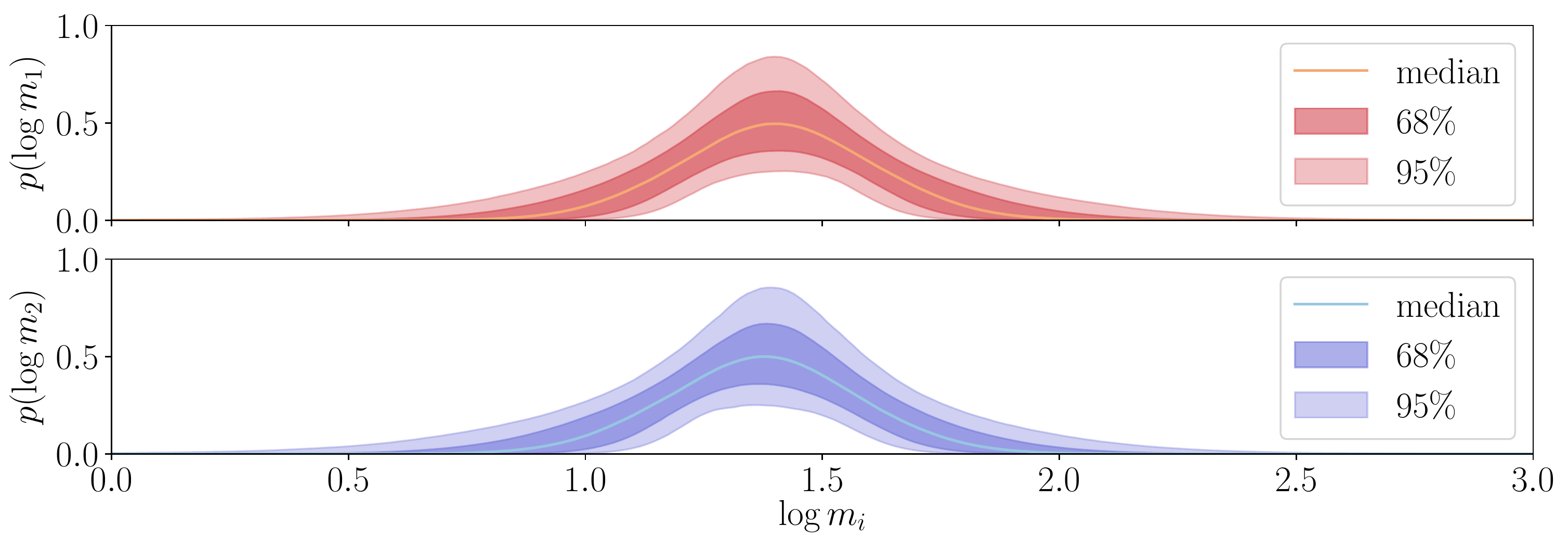}
\caption{\label{fig:pmi_2COMP}
The recovered marginal mass distributions on the \emph{observed} component masses in source frame. 
The red (blue) curves show the marginal distribution on $\log m_1$ ($\log m_2$). All masses are measured in solar mass units.
The central line in each case corresponds to the a posteriori median values of $p(\log m_i \mid\mathbf{\Lambda})$. 
The shaded regions denote the $1\sigma$ and $2\sigma$ confidence regions associated.
These posteriors are obtained by marginalizing over $K$. The two dimensional mass distribution is shown in Fig.~\ref{fig:LIGO_2Ddistro} in the Supplementary material.
}
\end{figure*}

We model the observed mass distribution using $K=1,\dots,4$ Gaussian components.
We sample the distribution in Eq.~\ref{eq:pop_post} using the nested sampling algorithm \cite{2004AIPC..735..395S} 
as implemented in CPNest \cite{john_veitch_2017_825456}. 
The primary output of the algorithm is the model evidence, which we use to determine which $K$ is favored; we find that the data favors a description using a 1-component Gaussian mixture.
Additionally, the algorithm produces samples from the posterior in Eq.~\ref{eq:pop_post}.
The log-evidences for different $K$ are presented in  Table~\ref{tab:logZ_LIGO}, while the posterior samples for $K=2$ ($K=3$) on the $\vec{\mu}^{(\log m_1)}$ parameters are shown in Fig.~\ref{fig:mix_HT_HC} (Fig.~\ref{fig:LIGO_posterior} in Suppl.~\ref{supp:LIGOVirgo_pop}).
The full posterior chain on all of the parameters is provided at~\cite{supplementary}.

Because this is a relatively low-dimensional problem (we consider $K\leq 4$ ) the analysis can be performed both with and without the hypertriangle map.
If the map is not used then the posterior has $K!$ degenerate modes.
If the map is used then there is just a single mode and, importantly, no information is lost.
The elimination of the excess multimodality is shown for two dimensions in Fig.\ref{fig:mix_HT_HC}.
More impressive demonstrations of the elimination of the excess multimodality are possible in higher numbers of dimensions; a plot in 3 dimensions for $K=3$ component mixture is shown in the supplementary material.
The preservation of information is demonstrated by the fact that the evidence in unchanged.
That the evidence is unchanged can be shown analytically and is a consequence of the Jacobian for our transformation in Eq.~\ref{eq:Jac_phi}; it is also demonstrated numerically for this specific problem in Table~\ref{tab:logZ_LIGO}.

We can now use the posteriors on $\mathbf{\Lambda}$ to plot the observed black hole mass distribution. 
This can be done using the posterior on the $\mathbf{\Lambda}$ from either of Eqs.~\ref{eq:LIGO_comp} or \ref{eq:LIGO_comp_2} with identical results. Although, the single mode posterior from Eq.~\ref{eq:LIGO_comp_2} is naturally easier to sample from.
The marginalised mass distributions on $m_1$ and $m_2$ are plotted in Fig.~\ref{fig:pmi_2COMP}.
As shown by the evidences in Table \ref{tab:logZ_LIGO} a one component mass distribution is favoured. 
We stress again that we have not included selection effects; including these is expected to suppress the high mass tail (this is because high mass BBHs can be seen out to greater distances than lower mass systems) and therefore our results are not incompatible with the presence of a mass gap.

The hypertriangle map has demonstrated its utility. 
It eliminated the excess multimodality in the description of the observed BBH mass distribution. This renders the target posterior easier to sample. There is no loss in information incurred by sampling this remapped parameter space compared to sampling the full original space.

\subsection{Overlapping Galactic White-Dwarf \\ Binaries in LISA}\label{subsec:4b}

LISA \cite{2017arXiv170200786A} is a planned space-based mission which will observe GWs in the (0.1--100)~mHz frequency range.
The LISA band is source-rich, with many signals overlapping in both time and frequency.
In particular, galactic white dwarf binaries (GBs) \cite{2017JPhCS.840a2024C} are so numerous at low frequencies that they form a confusion noise foreground for LISA.
Several GBs have already been identified electromagnetically and will serve as verification sources for LISA \cite{2006CQGra..23S.809S}.

The label-switching problem arises in the analysis of multiple sources, since the parameters of any pair of sources are interchangeable.
In this section we will show how the application of the hypertriangle map allows for efficient Bayesian recovery of multiple GB signals without ambiguity arising from label switching.

The GWs emitted by a distant source are observed in the solar system as plane waves.
There are two GW polarization components denoted $+$ and $\times$.
Under the assumption that each source is monochromatic, these components are given by
\begin{align}
    h_+(t; \mathbf{\Lambda})      &= A\,(1+\cos^2\iota)\,\cos(2\pi ft-\Phi) \,,\nonumber\\
    h_\times(t; \mathbf{\Lambda}) &= -2A\,\cos\iota\,\sin(2\pi ft-\Phi) \,,
    \label{eq:GB_h+_hx}
\end{align}
where $f$ is the GW frequency, $\iota$ is the inclination angle between the binary's orbital angular momentum and the line of sight, and $\Phi$ is a phase offset. 

The LISA detector response additionally depends on the ecliptic longitude and latitude $\{\lambda, \beta \}$ of the source and a polarization angle $\psi$.
The GW amplitude $A$ can be further expressed in terms of physical quantities of the GB system (e.g.\ the component masses and the luminosity distance); however, these quantities are highly degenerate and are therefore not considered.

Each of the $K$ sources is described by seven parameters:
\begin{align}
    \mathbf{\Lambda}_k = \{\log_{10} A_k, f_k, \lambda_k, \sin \beta_k, \cos\iota_k, \psi_k, \Phi_k\} \,.
\end{align}
We use flat priors on all parameters with ranges given in  Sec.~\ref{subsec:S3}.
The log-likelihood is given by
\begin{align}
    \log\mathcal{L}(\mathbf{\Lambda}_k) \propto -\frac{1}{2}\sum_{\alpha}\left| s_\alpha - \sum_{k=1}^{K} h_\alpha(\mathbf{\Lambda}_k)\right|_{(\alpha)}^{2}\,,
    \label{eq:lisalike}
\end{align}
where $k$ labels the various GBs, and where $s_\alpha$ denotes two approximately independent LISA output channels, with $\alpha\in\{\mathrm{A},\mathrm{E}\}$ (see, for example, \cite{2014LRR....17....6T}).
The model $h_\alpha$ is the LISA response to sinusoidal signals of the form in Eq.~\ref{eq:GB_h+_hx}.
The line brackets indicate a norm with respect to the usual signal inner product
\begin{align}\label{eq:inner_prod}
    \left<a|b\right>_{(\alpha)} = 4\Re\left\{\int_{0}^{\infty}\mathrm{d}f\; \frac{\tilde{a}(f)\tilde{b}(f)}{S_{\alpha}(f)}\right\} \, ,
\end{align}
where $\tilde{a}(f)$ is the Fourier transform of $a(t)$. 
Each output channel is assumed to contain additive stationary Gaussian noise with a one-sided power spectral density $S_\alpha(f)$.

    \begin{table}
        \begin{center}
            \begin{tabular}{cccccccc}
                $f\!-\!f_\star $ & $\log_{10} A\;$ & $\iota\,$[rad]$\;$ & $\lambda$ [rad]$\;$ & $\beta\,$[rad]$\;$ & $\psi\,$[rad]$\;$ & $\phi\,$[rad]$\;$ & $\rho$ \\
                \hline
                $0$ & -22.15& 0.246& -0.096& 0.218& 1.640& 1.795& 10 \\
                $2/\mathrm{yr}$ & -22.13& 0.403& 0.091& 0.294& 1.066& 4.249& 10 \\
                $4/\mathrm{yr}$ & -22.13& 0.376& -0.055& 0.359& 0.794& 4.760& 10 \\
                $6/\mathrm{yr}$ & -22.15& 0.284& 0.031& 0.248& 1.127& 2.078& 10 \\
                $8/\mathrm{yr}$ & -22.13& 0.390& 0.006& 0.223& 0.775& 4.537& 10 \\
                $10/\mathrm{yr}$ & -22.12& 0.428& 0.091& 0.296& 1.088& 5.765& 10 \\
            \end{tabular}
        \end{center}
        \caption{\label{tab:6WD} The parameters of the six injected GBs, with $f_\star =1\,\mathrm{mHz}$. The amplitudes were chosen such that the signal-to-noise ratio is 10 in each case.}
    \end{table}

We simulate one year of mock LISA noise using LISA code \cite{2008PhRvD..77b3002P}. For simplicity, we estimate the power spectral densities from these signal-free noise realizations using the Welch periodogram \cite{welch,bartlett}.

\begin{figure*}
    \centering
    \includegraphics[width=2\columnwidth]{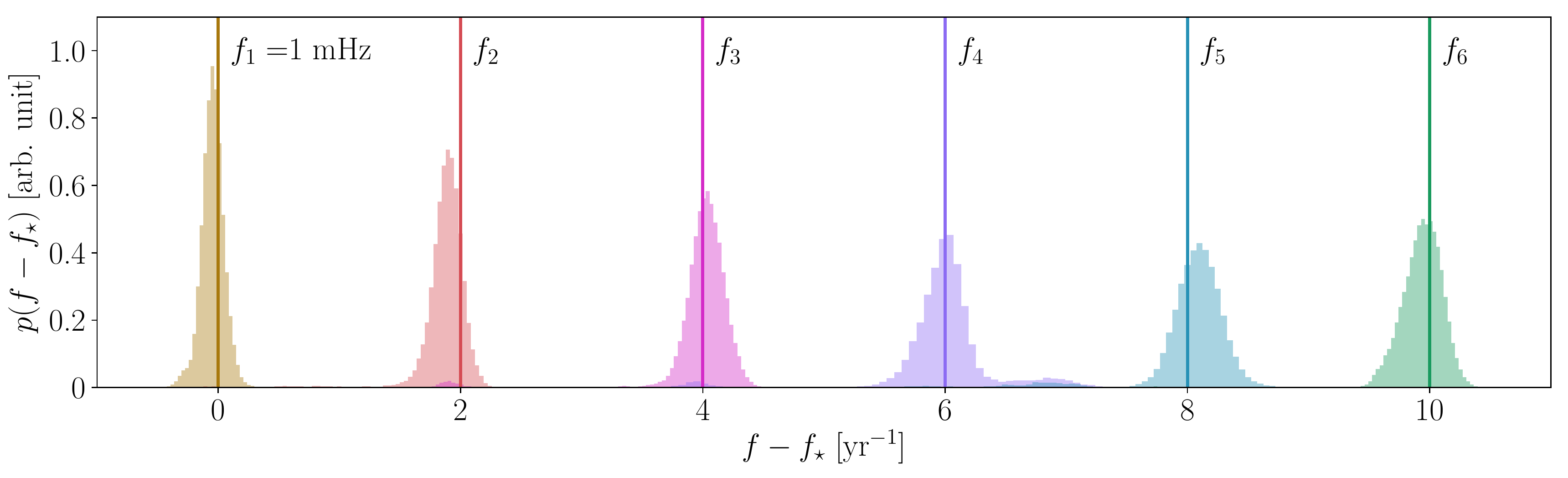}
    \caption{\label{fig:6WD}
        The 1-D marginalized posterior distributions on the physical frequencies $f_k$ of the six GBs. Vertical lines mark the injected frequencies. As we used a zero noise injection, we expect the posteriors to be peaked at the injected values.
        We observe that some neighboring sources (notably 4 and 5) show some correlation (see also the supplementary figures).  This effect is not an artifact of the hypertriangle map or the sampling.  Rather, it is a genuine feature of the posterior caused by the non-zero overlap between sources closely spaced in frequency.
    }
    \label{fig:LISA_6WD}
\end{figure*}

We inject $K=6$ sources, each with a signal-to-noise ratio $\rho_k=10$ defined with respect to the inner product in Eq.~\ref{eq:inner_prod}.
The six sources were chosen to have regularly spaced frequencies;
other source parameters were chosen randomly and are given in Table~\ref{tab:6WD}.
For simplicity, we perform a noise-free analysis.

The simulated data has a cadence of $5\,\mathrm{s}$ and a total duration of $1\,\mathrm{yr}$, resulting in arrays of length $6.3\times 10^6$.
This data was heterodyned, filtered, and downsampled to isolate a narrow range of frequencies $f\in(f_\star -1/\mathrm{yr}, f_\star +11/\mathrm{yr})$, where $f_\star  = 1$~mHz. 
For a one-year observation period, the expected frequency resolution of LISA is $\sim 1/1\,\text{yr}$, so this frequency range covers 12 bins.

We assume the number of sources $K$ is already known by other means; we do not address the problem of searching for an unknown number of sources (see, for example, \cite{2009PhRvD..80f4032S, 2007CQGra..24S.575C, PhysRevD.67.103001}).

This is a $6\times7=42$-dimensional Bayesian inference problem.
The likelihood in Eq.~\eqref{eq:lisalike} is invariant under permutations of the index $k$ (i.e.\ relabelling the GBs numbered 1 to 6).
Naively sampling this distribution in the specified prior ranges will return a posterior distribution with (at least) $6!=720$ peaks.
To remove this problem we enforce the artificial identifiability constraint $f_{k+1} \geq f_k$ by sampling on the parameters 
\begin{align}
    \mathbf{\Lambda}_k = \{\log_{10} A_k, \chi_k, \cos\iota_k, \lambda_k, \sin \beta_k, \psi_k, \phi_k\} \, .
\end{align}
Here $f_k = \phi(\chi_k)$ (see Eq.~\ref{eq:map2}), and the prior on $\chi_k$ is the same as the prior on $f_k$.
The sampler explores the space of $\chi_k \in \mathcal{C}$ which is mapped to the physical frequencies $f_k \in \mathcal{T}$.

The resultant distribution has a single global maximum and is therefore relatively easy to sample from (albeit in 42 dimensions).  
We use CPNest \cite{john_veitch_2017_825456} to sample the distribution and correctly recover all sources.  
We note that without applying our hypertriangle map, it would be excessively difficult to sample from this 720-fold degenerate distribution.

The 1D marginalized posteriors on the physical frequencies $f_k$ are plotted in Fig.~\ref{fig:6WD}. Additional plots and the full posterior file are provided in the supplementary material.  
\section{Discussion}\label{sec:5}

We discussed a general solution to the label switching problem which allows the sampler to be treated as a black box, and is therefore widely applicable. To enforce the identifiability constraint, we map the sampled points from a hypercube with the desired prior to a hypertriangle, taking care to preserve the prior. We have successfully used this for two real-world problems from gravitational wave astrophysics. 
The hypertriangle transformation has the potential to greatly simplify a wide class of highly-degenerate Bayesian inference problems, with no loss of information.

\acknowledgements

The authors gratefully acknowledge the contributions of  Alberto Vecchio, Sebastian Gaebel, Antoine Klein, Davide Gerosa, Alberto Sesana, Christopher Berry and Siyuan Chen in the development of the Birmingham LISA data analysis code which was used for the GB problem; 
full details of this code will be published elsewhere.
The authors also thank Alvin Chua, John Veitch and Walter Del Pozzo for discussions about the method and RB acknowledges Gianmarco Brocchi for useful discussions.
The computations described in this paper were performed using the University of Birmingham's BlueBEAR HPC service.

\bibliography{refs.bib}

\clearpage

\onecolumngrid

\renewcommand{\theequation}{S\arabic{equation}}  
\renewcommand{\thepage}{S\arabic{page}}  
\renewcommand{\thesection}{S\arabic{section}}   
\renewcommand{\thetable}{S\arabic{table}}   
\renewcommand{\thefigure}{S\arabic{figure}}

\setcounter{equation}{0}
\setcounter{page}{1}
\setcounter{section}{0}
\setcounter{table}{0}
\setcounter{figure}{0}

\center{
\textbf{
SUPPLEMENTARY MATERIAL
} 
}
\flushleft

\section{Additional LIGO/Virgo population Data and Plots}\label{supp:LIGOVirgo_pop} 

In this section, we show prior and posteriors for parameters of the observed populations described in Sec. \ref{subsec:4a}. 

\begin{figure}[ht]
    \centering
    \includegraphics[width=\textwidth]{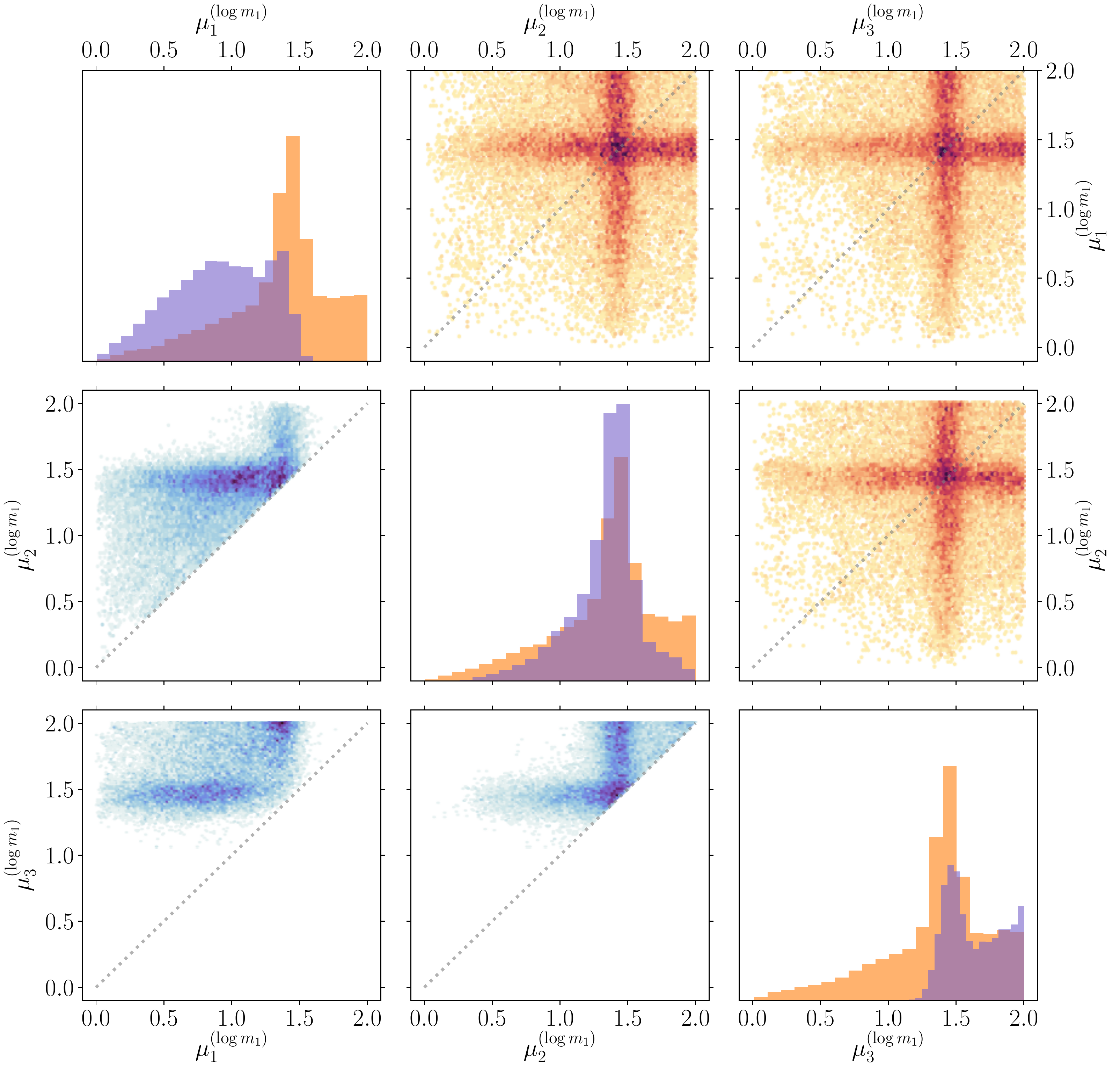}
    \caption{\label{fig:LIGO_posterior}Posteriors on mixture parameters  $\left(\mu_1^{(\log m_i)},\mu_2^{(\log m_i)},\mu_3^{(\log m_i)}\right)$, assuming $K=3$. 
    The bottom-left and top-right triangles show the corner plots for the analysis performed the with and without applying the hypertriangulation map, respectively hypertriangulation map. Along the diagonal the 1-dimensional samples histograms are overlaid with both configurations. Dashed gray lines denote equal mixture components primary mass means.}
\end{figure}

\begin{figure}[ht]
    \centering
    \includegraphics[width=0.5\columnwidth]{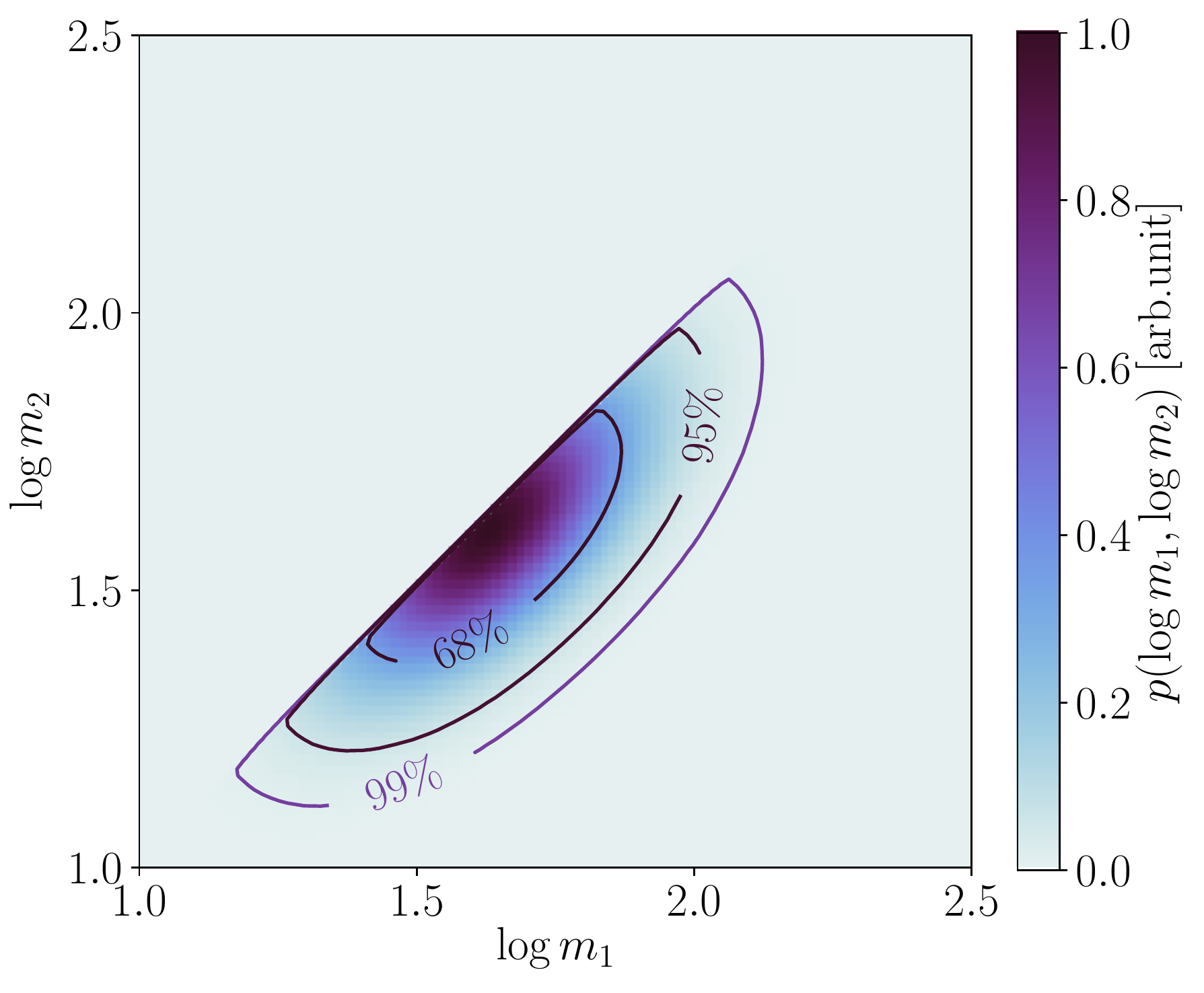}
    \caption{\label{fig:LIGO_2Ddistro}$p(\log{m_1},\log{m_2})$ a posteriori median values. Posterior samples from the one and two component mixtures are combined, according to their evidence, into a single set of posterior samples..  Lines denote the $1\sigma$, $2\sigma$ and $3\sigma$ contour levels, respectively.} 
    
\end{figure}

Priors on mixture parameters are given in Table \ref{tab:LIGO_priors}. Posteriors on mixture primary log-mass means are shown in Fig.~\ref{fig:LIGO_posterior}. Median a posteriori values of $p(\log m_1,\log m_2)$ are shown in Fig.~\ref{fig:LIGO_2Ddistro} for one and two mixture components.

\begin{table}[ht]
    \begin{center}
        \begin{tabular}{cccccc}
            $\mu_k^{(\log m_1)}$  & $\mu_k^{(\log m_1)}$ & $\lambda_k^1$ & $\lambda_k^2$ & $\phi_k$ [rad] & $w_k$\\
            \hline
            $\quad\;$(0, 2)$\quad\;$ & $\quad\;$(0, 2)$\quad\;$ & $\quad\;$(0.01, 4)$\quad\;$ & $\quad\;$(0.01, 4)$\quad\;$ & $\quad\;$(0, $\pi/2$)$\quad\;$ & $\quad\;\;$(0, 1)$\quad\;\;$
        \end{tabular}
    \end{center}
\caption{\label{tab:LIGO_priors} 
Priors on the BBH observed population parameters.}
\end{table}
\section{Additional GB Data and Plots}
\label{subsec:S3}
\begin{figure}
    \centering
    \includegraphics[width=\columnwidth]{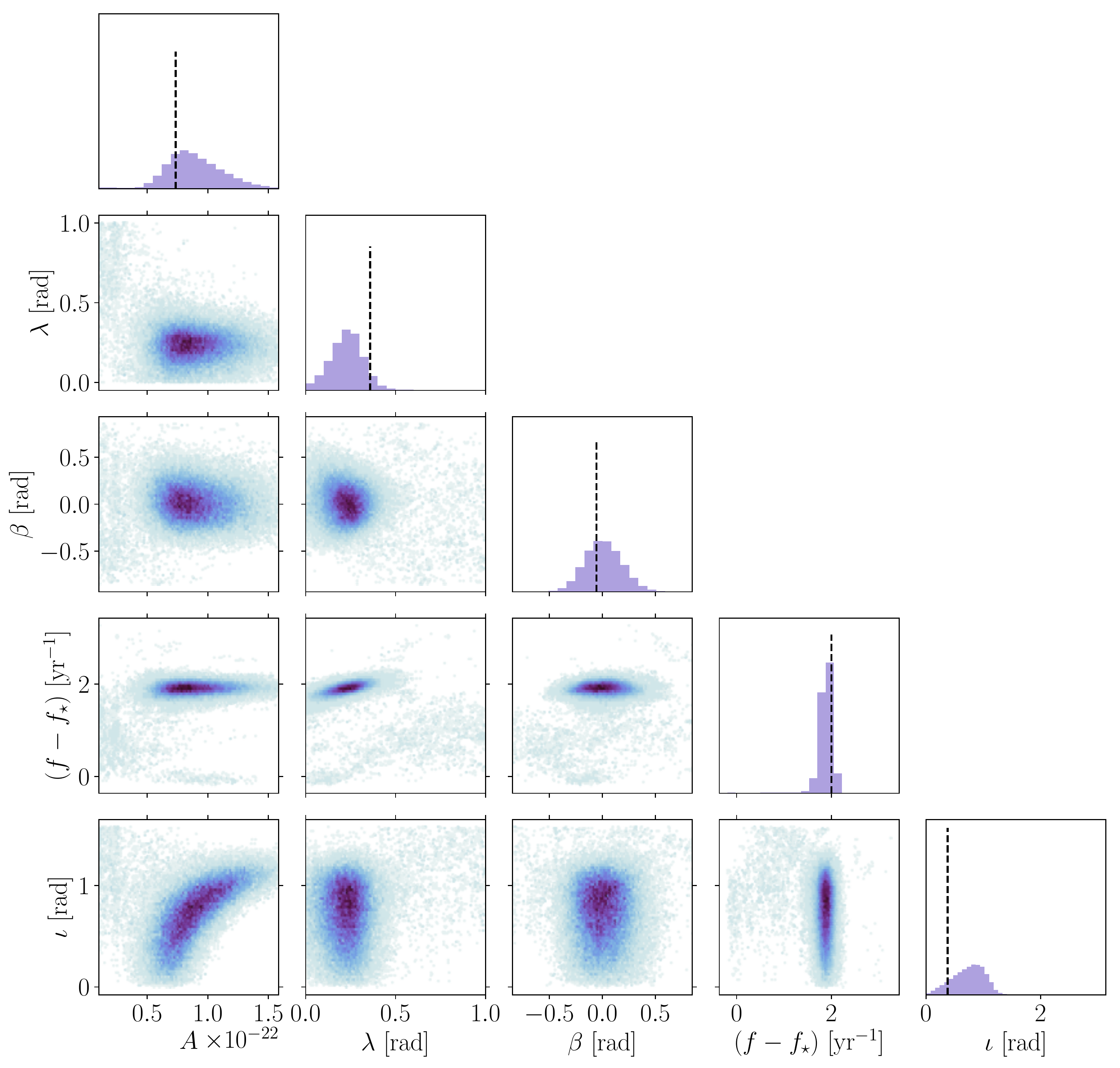}
    \caption{\label{fig:selection} Posteriors on selected parameters from the third galactic binary. Vertical lines show the true injected values.}
\end{figure}

The full posterior parameters are in \cite{supplementary}. A selection of these parameters are plotted in Fig.~\ref{fig:selection}
\begin{table}[ht]
        \begin{center}
            \begin{tabular}{ccccccc}
                 $(f - f_\star )$ $[\text{yr}^{-1}]$  & $\log_{10} A$ & $\lambda$ [rad] & $\sin \beta$ & $\cos \iota$ & $\psi$ [rad] & $\Phi$ [rad]  \\
                \hline
                 $\quad\;\;$(-1, 11)$\quad\;\;$&  $\quad$(-23.0, -21.8)$\quad$ & $\quad$(0, 1)$\quad$ & $\quad$(-0.75, 0.75)$\quad$ & $\quad$(0,1)$\quad$ & $\quad$(0, $\pi$)$\quad$ & $\quad$(0, $2\pi$)$\quad$
                \end{tabular}
        \end{center}
        \caption{\label{tab:GB_priors} 
        Priors on the GB parameters.
        The frequency prior spans twelve bins around $f_\star =1\,\mathrm{mHz}$.
        }
    \end{table}

\end{document}